\documentclass[manuscript]{emulateapj}
\usepackage{natbib,times}
\usepackage{lscape}

\shorttitle{Stellar populations of ULIRGs}
\shortauthors{ L.G. Hou et al.}
\begin{document}

\title{Stellar populations of ultraluminous infrared galaxies}

\author{L.~G. Hou$^1$, J.~L. Han$^1$,  M.~Z. Kong$^2$, Xue-Bing Wu$^3$}

\altaffiltext{1}{National Astronomical Observatories, Chinese Academy of Sciences,
                 Jia-20 DaTun Road, Chaoyang District, Beijing 100012, China}
\altaffiltext{2}{Department of Physics, Hebei Normal University,
                 Shijiazhuang 050016, China}
\altaffiltext{3}{Department of Astronomy, School of Physics, Peking University,
                 Beijing 100871, China}

\begin{abstract}
Ultraluminous infrared galaxies (ULIRGs) have several types according
to dominance of starburst or AGN component. We made stellar population
analysis for a sample of 160 ULIRGs to study the evolution of
ULIRGs. We found that the dominance of intermediate-age and old
stellar populations increases along the sequence of HII-like ULIRGs,
Seyfert-HII composite ULIRGs, and Seyfert 2 ULIRGs. Consequently the
typical mean stellar age and the stellar mass increase along the
sequence. Comparing the gas mass estimated from the CO measurements
with the stellar mass estimated from the optical spectra, we found
that gas fraction is anti-correlated with the stellar mass. HII-like
ULIRGs with small stellar masses do not possess enough gas and the
total mass, and therefore have no evolution connections with massive
Seyfert 2 ULIRGs. Only massive ULIRGs may follow the evolution
sequence toward {AGNs}, and massive HII-like ULIRGs are probably in an
earlier stage of the sequence.
\end{abstract}
\keywords{galaxies: infrared --- galaxies: evolution
  --- galaxies: starburst --- galaxies: stellar content}

\section{Introduction} \label{sect:intro}

Ultraluminous infrared galaxies (ULIRGs) are advanced mergers of
gas-rich galaxies. They have an infrared luminosity greater than
10$^{12}$~L$_\odot$ in 8-1000 $\mu$m band \citep[e.g.][]{sm96,lfs06}.
Their high infrared luminosity comes from violent starbursts with
significant contribution from active galactic nuclei (AGNs) in some
ULIRGs \citep[e.g.][]{fae03}.

ULIRGs can be classified according to dominance of starburst or AGN
component, namely as the HII-like ULIRGs, LINER ULIRGs, Seyfert-HII
composite ULIRGs, Type 1 ULIRGs and Seyfert 2
ULIRGs. \citet{sse+98,sse+982} proposed an evolutionary scenario of
ULIRGs from cool-ULIRGs to warm-ULIRGs and finally to QSOs. The
cool-ULIRGs have $f_{25}/f_{60} \le 0.2$ and are preferably
star-forming galaxies \citep{ham87}, i.e. the HII-like galaxies. Here
$f_{25}$ and $f_{60}$ are flux densities at 25~$\mu m$ and 60~$\mu m$,
respectively. The warm-ULIRGs have $f_{25}/f_{60} > 0.2$ and
preferably host an AGN \citep{dmlj85,bsh+86}. ULIRGs are important to
understand the evolution of merger galaxies and the possible
connections between the starburst and AGNs.  Further imaging and
spectroscopy studies
\citep[e.g.][]{ssv+98,ss99,sse00,ltd+03,sls+06,nls+07,vrk+09} and
numerical simulations \citep[e.g.][]{ks92,sdh05,njb06} support this
evolutionary scenario. However, this standard evolution sequence is
questioned by \citet{cbb+01,gtr+01,tgl+02,rtg10}.

If the standard evolutionary scenario is true, ULIRGs should
experience the continuing conversion of gas component to stellar
populations in {their} lifetime. ULIRGs in the earlier evolutionary
stage are expected to possess more gas and young stars, less old
stellar populations, smaller stellar mass and larger gas mass fraction,
compared with those of ULIRGs in a later evolutionary stage. There
should be enough gas in early stage ULIRGs to be converted into
stars when they evolve to a later stage for a larger stellar
mass.

The stellar population analysis of ULIRGs can shed light on ULIRG
evolution. Previous such analyses were only carried out on very small
samples \citep[e.g.][]{cs00,cs002,cs01,rhtg07,rtg08,sm10}. A slightly
larger sample of 36 local ULIRGs was studied by
\citet{rtg09,rtg10}. In contrast to the standard evolutionary
scenario, their stellar population analysis based on the optical
spectra exhibits no significant differences of the stellar ages among the
HII-like ULIRGs, LINER, and Seyfert 2 ULIRGs. Note that {the} physical
parameters of ULIRGs are rather scatter \citep[e.g.][]{vrk+09},
{therefore,} the possible evolutionary of ULIRGs can not be revealed
if only a small sample is applied to.

In this work, we {analyse} the spectra of 160 ULIRGs available from
the Sloan Digital Sky Survey \citep[SDSS,][]{aaa+09} for stellar
populations. In Section 2, we introduce our ULIRG sample. The stellar
population analysis of the sample is presented in Section 3.  In
Section 4, we discuss the inferred stellar age, mass and gas for
different types of ULIRGs. {Several} factors which affect our analysis
are discussed in Section 5, and conclusions are given in Section 6.

In this paper, we adopt $H_0$ = 70 km s$^{-1}$ Mpc$^{-1}$,
$\Omega_{\rm m} = 0.3$ and $\Omega_{\rm \Lambda} = 0.7$.


\section{The ULIRG sample}
\label{sect:Obs}

For reliable stellar population analysis, we work on a large sample
of ULIRGs which have optical spectra
covering the wavelength regions of 4000~$\rm \AA$ break and strong
stellar absorptions near 4000~$\rm \AA$.

\begin{figure*}[bt]\centering
\includegraphics[angle=270,width=56mm]{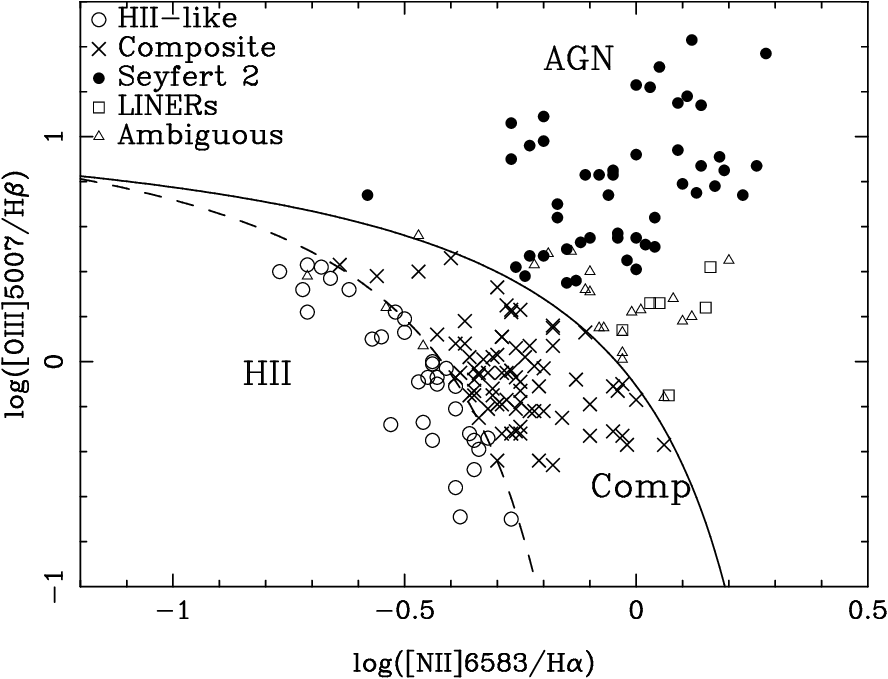}
\includegraphics[angle=270,width=56mm]{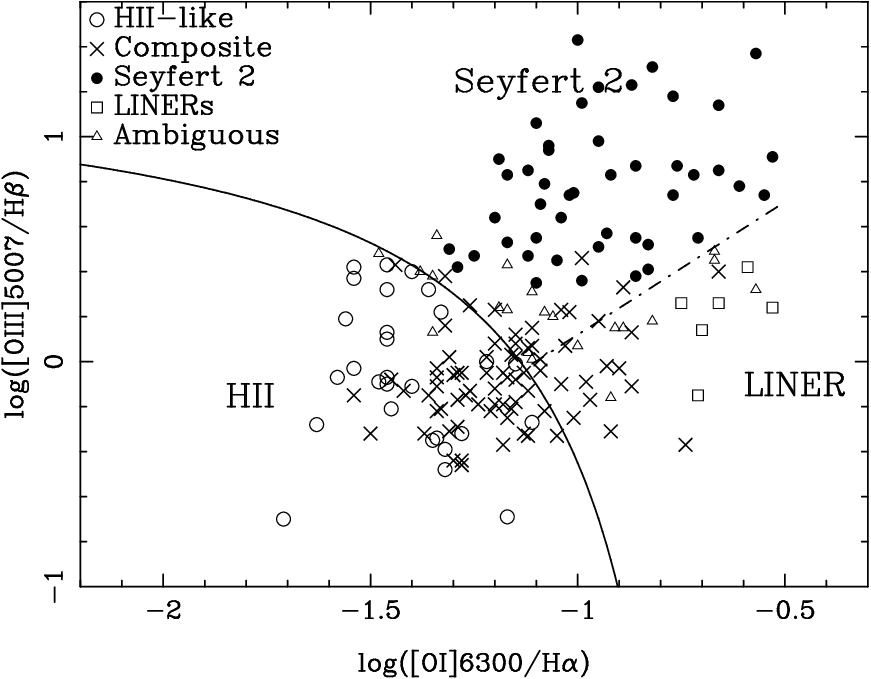}
\includegraphics[angle=270,width=56mm]{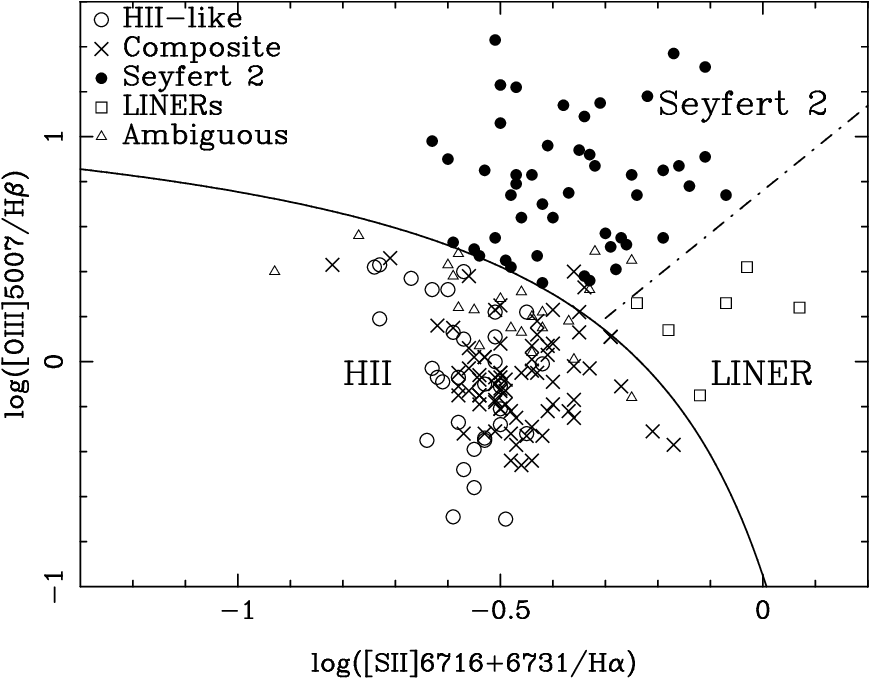}
\caption{BPT diagrams for line ratios of 185 narrow-line ULIRGs.}
\label{bpt}
\end{figure*}

\begin{figure}[bt]
\centering \includegraphics[angle=270,width=83mm]{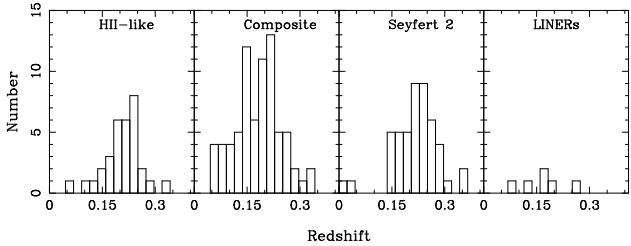}
\caption{Redshift distribution of 160 narrow line ULIRGs, after we
  discarded 22 ambiguous objects and 3 Seyfert-HII composite ULIRGs
  with bad pixels from 185 narrow-line ULIRGs.}
\label{disz}
\end{figure}

The SDSS database \citep[DR7,][]{aaa+09} released the spectra of more
than 920,000 galaxies in the range of 3800~$\rm \AA$ to 9200~$\rm \AA$
in the observer frame. Following the procedures described in
\citet{hwh09}, we found 398 ULIRG candidates from the SDSS DR7
spectral catalog \citep{aaa+09} and {\it IRAS} Faint Sources Catalog
\citep{mkc92}. We checked them from NASA/IPAC Extragalactic Database
(NED) and SDSS images. We found that 38 of them are wrong
identifications of ULIRGs due to either the known redshift of a ULIRG
in the NED not consistent with the redshift from the SDSS database or
an obviously nearby large bright galaxy as an IRAS source but with a
mis-matched SDSS redshift. The spectra of 360 identified ULIRGs are
then fitted and classified \citep{hwh09}. Among them, 73 ULIRGs have
broad lines and are of Type 1, and 287 objects are narrow-line ULIRGs.
We will not work on Type 1 ULIRGs for stellar population studies in
this paper because their spectra are dominated by strong AGN features
(e.g. the power-law continuum, the strong and broad emission lines,
the strong {  Fe$_{\rm II}$} features).

Based on the line ratios in {the} BPT diagrams \citep[see
  Figure~\ref{bpt},][]{bpt81,kgkh06}, 185 of 287 narrow-line ULIRGs
are further classified as 32 HII-like ULIRGs, 77 Seyfert-HII composite
ULIRGs, 48 Seyfert 2 ULIRGs, 6 LINER ULIRGs, and 22 ambiguous
objects. We omitted the other 102 narrow-line ULIRGs from stellar
population analysis, because one or more lines either are redshifted
outside the wavelength range or do not have a good signal-to-noise
ratio for the classification in the BPT diagrams. We also discarded 22
ambiguous ULIRGs for our analysis because they are shown as one type
in {one} BPT {diagram} but another type in the other diagram(s). Three
Seyfert-HII composite ULIRGs are further skipped because more than
half of pixels in their SDSS spectra are masked as bad. Therefore, we
will study the stellar populations of 160 well-classified
ULIRGs. Their redshift distribution is shown in Figure~\ref{disz}.

\begin{figure}[tb]
\centering \includegraphics[angle=270,width=85mm]{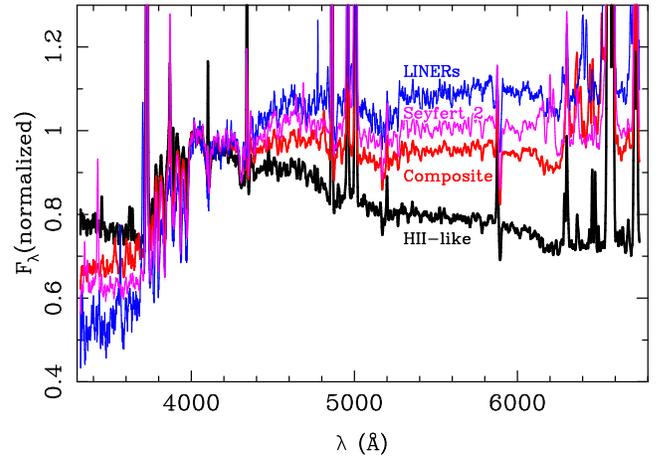}
\caption{Combined spectra of {  four types} of narrow-line
ULIRGs.} \label{combin}
\end{figure}

In Figure~\ref{combin} we show the combined spectra for four types of
narrow-line ULIRGs. The spectrum of every ULIRG is first normalized by
the {  median} flux in the wavelength $4010~{\rm \AA}$ -- $4060~{\rm
 \AA}$. The combined spectra are the weighted mean of the normalized
spectra according to their signal-to-noise ratio in the wavelength
range of $5300~{\rm\AA}$ --$5800~{\rm\AA}$ in the rest frame. The
combined spectra clearly show that along the sequence of HII-like,
Seyfert-HII composite, Seyfert 2, and LINER ULIRGs, the normalized
continuum fluxes below 4000 ${\rm \AA}$ decrease and the fluxes above
4000 ${\rm \AA}$ increase, which roughly indicates that the fraction
of old stellar populations increases along the sequence.

\section{Stellar population analysis}

The stellar population analysis is an important tool to study star
formation in galaxies and evolution of galaxies. It has been applied
to various type of galaxies, e.g. HII galaxies \citep[e.g.][]{sbp96,
wctk04}, AGNs \citep[e.g.][]{cgs+04,cgm+04}, infrared selected
galaxies \citep{clh+09,clh+10}, ULIRGs \citep[e.g.][]{rtg10,mwg+10}.
Here, we use it for a large ULIRG sample to study the evolution
sequence of ULIRGs.

We noticed that the fiber diameter of SDSS spectrograph is $3''$,
corresponding to $\sim$~3~kpc for a galaxy at redshift $z\sim0.05$
or $\sim18$~kpc at $z\sim0.25$. For any galaxies with different
redshifts, the observed spectrum comes from central regions of
different size. We use the software,
STARLIGHT \footnote{http://www.starlight.ufsc.br} \citep[{  version
04, }][]{cms+05,msc+06,acs+07}, to analyze the stellar populations of
ULIRGs in the central {  regions}. We will discuss the aperture
effect later.

\begin{figure*}[bth]
\centering
\includegraphics[angle=270,width=140mm]{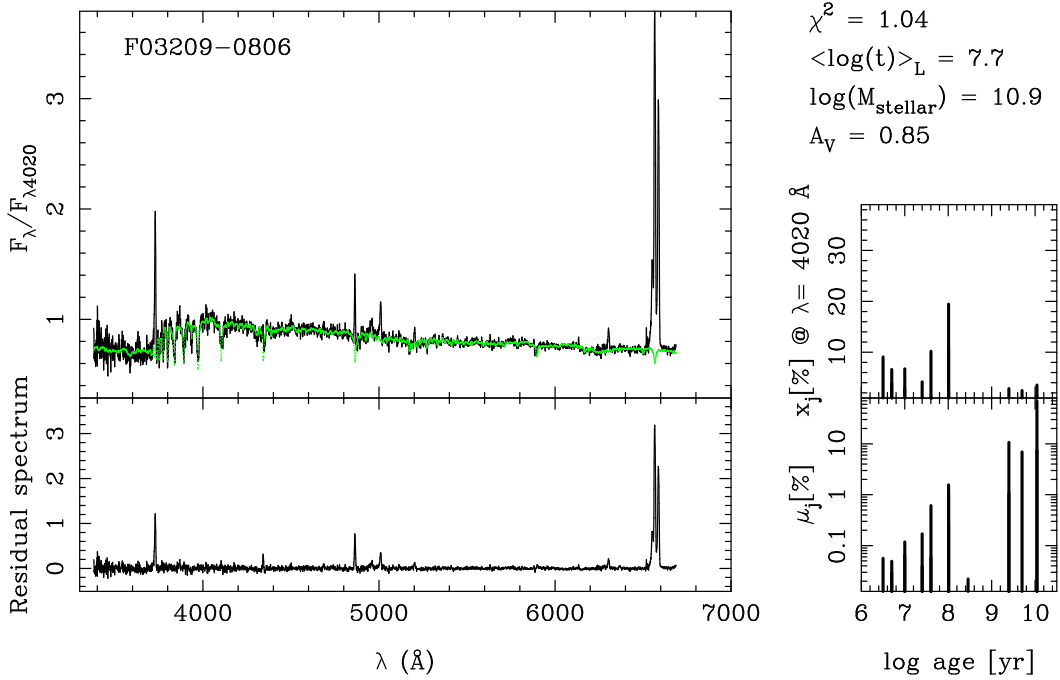}
\caption{{  One example} for spectrum fitting. We show the
  normalized observed spectrum ($F_{\lambda}/F_{\lambda4020}$, black)
  and the modeled spectrum (green) {  in the left-upper panel and
  the residual in the left-lower panel}. The key parameters from the
  fitting: $\chi^2$, mean stellar age, stellar mass and extinction
  parameter are given in the right side. The fractional
  contribution to the model flux ($x_j$) and the stellar mass
  ($\mu_j$) of the $j$th SSP are plotted in the right.}
\label{2examp}
\end{figure*}

The STARLIGHT fits an observed galaxy spectrum $O(\lambda)$ with a
combination of $N$ simple stellar populations (SSPs). The intrinsic
extinction due to the foreground dust in the host galaxy is
considered and parameterized by the $V$-band extinction $A_V$. The
line-of-sight stellar motions are modeled by a Gaussian distribution
$G(v_0,\sigma_*)$ centered at the velocity $v_0$ and with the
dispersion $\sigma_*$. The modeled spectrum $M(\lambda)$ is
described by:
\begin{equation}
M(\lambda)=M_{\lambda_0}[\sum_{j=1}^{N}x_jb_{j}(\lambda)\otimes
G(v_0,\sigma_*)]r({\lambda})
\end{equation}
where $M_{\lambda_0}$ is the synthetic flux normalized at
$\lambda_0$, $x_j$ is the fractional contribution of the $j$th SSP
to the model flux at $\lambda_0$, $b_{j}(\lambda)$ is the normalized
spectrum of the $j$th SSP, and
$r(\lambda)=10^{-0.4[A(\lambda)-A_{\lambda_0}]}$ is the reddening
term. The best fitting to search for the minimum
$\chi^2=\sum_{\lambda}[(O(\lambda)-M(\lambda))\cdot w(\lambda)]^2$,
where {  $w(\lambda)$ is the weight factor, and $w(\lambda)^{-1}$ is
  the uncertainty of observed spectrum $O(\lambda)$ given in the
  SDSS database}.
We used a base of $N=45$ SSPs with different ages and metallicities
from the evolution synthesis models of \citet{bc03}, and adopted the
initial mass function from \citet{cha03} and the Padova 1994
evolutionary tracks \citep{abb+93,bfbc93,fbbc94,gbc+96} and STELIB
library \citep{lbp+03}. The base comprises of star population with 15
different ages between 1~Myr to 13~Gyr at each of the three
metallicities: 0.2, 1, and 2.5 Z$_\odot$. Here Z$_\odot$ is the
metallicities of the Sun. The reddening law of \citet{cks94} is
adopted in the fitting.

We obtained the spectra with uncertainties of 160 ULIRGs from SDSS.
We corrected the Galactic reddening effect, and converted each
spectrum to the rest-frame. The STARLIGHT was used to fit the
continuum and absorption features. The emission lines and sky lines
are discarded in the mask file of STARLIGHT. The spectrum pixels
without error measurements or with negative flux values are
excluded. In addition, the Na D doublet, $5870~{\rm \AA}-5905~{\rm
  \AA}$, and other three bands, $6845~{\rm \AA}-6945~{\rm \AA}$,
$7550~{\rm \AA}-7725~{\rm \AA}$, and $7165~{\rm \AA}-7210~{\rm \AA}$
are also masked, due to the bugs in SSP model  {  \citep{lbp+03}}
or the large fitting residual in some regions \citep{msc+06}.
Considering the redshift coverage of our sample, we restricted the
fitting in the wavelength range $3400~{\rm \AA}-6700~{\rm \AA}$ for
each spectrum. A power-law component,
$F(\lambda)\propto\lambda^{\alpha}$, is used in the STARLIGHT to
account the AGN contribution to the observed continuum
\citep{cgm+04, cms+05}, and $\alpha=-1.5$ \citep{rs80,cgm+04} is
adopted. The random Markov Chains method was used in the STARLIGHT,
which needs an input of integer seed. Different seeds slightly
affect the fitting results. Following \citet{mwg+10}, we fit each
spectrum 100 times with different seeds. The final fitted parameters
are the mean values from 100 runs. One example of the spectrum
fitting {  is} shown in Figure~\ref{2examp}.

\begin{deluxetable}{lcccc}[bth]
\tabletypesize{\scriptsize} \tablecolumns{5} \tablewidth{0pt}
\tablecaption{{  Mean fractional contributions of different stellar
  populations} and the power-law component to the model flux at 4020
  $\rm \AA$ for different types of ULIRGs}
\tablehead{
\colhead{Type} & \colhead{young} & \colhead{intermediate-age}
& \colhead{old} & \colhead{power-law} \\
 &($<10^8$ yr) & ($10^8-10^9$ yr)& ($>10^9$ yr) &
}
\startdata
HII-like  & 0.622  & 0.261  & 0.117 &     \\
Composite & 0.355  & 0.408  & 0.138 & 0.099\\
Syefert 2 & 0.177  & 0.500  & 0.198 & 0.125\\
LINER     & 0.112  & 0.579  & 0.261 & ~0.048
\enddata
\label{tab_age}
\end{deluxetable}

\begin{figure}[bth]
\centering
\includegraphics[angle=270,width=85mm]{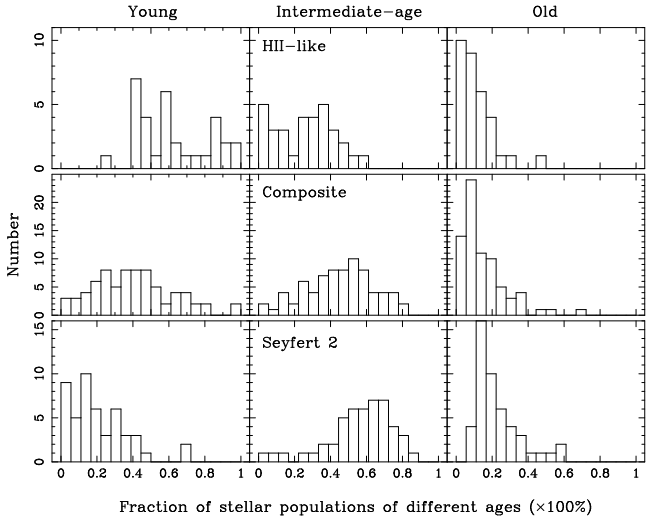}
\caption{Fraction distributions of stellar populations
for different types of ULIRGs from top to bottom.}
\label{frac}
\end{figure}

The stellar populations in a galaxy can be described by the
fractions of young stars of $t_j < 10^8$~yr, intermediate-age stars
of $10^8$~yr~$\le t_j \le 10^9$~yr, and old stars of $t_j > 10^9$~yr
\citep{cms+05}. Here $t_j$ is the age of the $j$th SSP.  We obtain
the fractional contributions of these three stellar populations of
different ages and the power-law contribution to the model { 
spectrum} flux at 4020 $\rm \AA$. For each type of ULIRGs, we got
their means as given in Table~\ref{tab_age}. Because there are only
a small number (only 6) of LINER ULIRGs, we will not discuss their
stellar population in the following.

The fractional contributions ({  i.e. $x_j$ in Equation~1}) of these
stellar populations to the total stellar emission are shown in
Figure~\ref{frac}. We found that they are very different for different
types of ULIRGs. The young stellar population is more dominant for the
HII-like ULIRGs than the composite and Seyfert 2 ULIRGs. The stellar
populations of intermediate-age and old stars are more dominant in the
Seyfert 2 ULIRGs than HII-like and composite ULIRGs. The older stellar
populations and the significant power-law component in Seyfert 2 ULIRGs
suggest that they are at a more evolved stage in the evolution track
towards AGNs. The fraction distribution of stellar populations of
composite ULIRGs {  suggests} that they are at the transitional
stage between HII-like and Seyfert 2 ULIRGs.

\section{Physical parameters of different types of ULIRGs}

The results of stellar population analysis of ULIRGs suggest the
possible evolution sequence from HII-like, composite, to Seyfert 2
ULIRGs. The HII-like ULIRGs are at an earlier evolutionary stage, the
composite ULIRGs are at a transitional stage, and the Seyfert 2 ULIRGs
are at a more evolved stage. In this section, we discuss the physical
parameters of ULIRG galaxies, namely, the mean stellar age, stellar
mass, and gas mass, which may be related to the possible
evolutionary sequence.

\subsection{The mean stellar age}

\begin{figure}[tb]
\centering \includegraphics[angle=270,width=85mm]{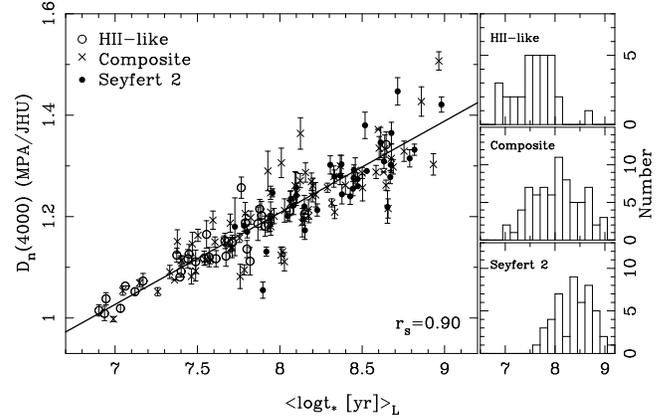}
\caption{Light-weighted mean stellar ages of ULIRGs estimated by
  the stellar population analysis are compared with the D$_n$(4000)
  obtained by the MPA/JHU group. The solid line is the best fit for
  all ULIRGs.  The distributions of the mean stellar ages for
  each type of ULIRGs are shown in the right panels.}
\label{age}
\end{figure}

Based on the results of the stellar population analysis, we
can get the light-weighted mean stellar age \citep{cms+05}:
\begin{equation}
\langle \log t_\ast  \rangle_L=\sum_{j=1}^{N} x_j \log t_j,
\end{equation}
which is an indicator of star formation history. The other such an
indicator is D$_n$(4000) obtained by the MPA/JHU
group\footnote{http://www.mpa-garching.mpg.de/SDSS/DR7/}, which is
defined as the average flux density ratio for the bands $3850~ {\rm
 \AA}-3950~ {\rm \AA}$ and $4000~{\rm \AA}-4100~ {\rm \AA}$
\citep{bru83,bmy+99}. We compared them (Figure~\ref{age}) and
found that they are well correlated, with the Spearman Rank-Order
Correlation Coefficient of $r_s=0.90$. The best fitting is
\begin{equation}
D_n(4000)=-(0.24\pm0.06) + (0.18\pm0.01)
\langle \log t_\ast  \rangle_L .
\end{equation}

The distributions of $\langle log t_\ast \rangle_L$ for HII-like,
composite and Seyfert 2 ULIRGs are shown in the right panels of
Figure~\ref{age}. The mean values of ${\langle log t_\ast \rangle_L}$
for three types of ULIRGs, HII-like, composite, and Seyfert 2, are
7.6, 8.0 and 8.4, with the standard deviations of 0.4, 0.5 and 0.4,
respectively. The increase of these mean values is consistent with the
evolution sequence from HII-like ULIRGs and composite ULIRGs to
Seyfert 2 ULIRGs. The young stellar population ($t_j<10^8$~yr) are
always present in ULIRGs (see Figure~\ref{age}) due to the enhanced star
formation in galaxy merger process. The large deviations of $\langle
log t_\ast \rangle_L$ for these types of ULIRGs are probably caused by
the complex dynamical process and star formation history of merger
systems. Therefore, both $\langle log t_\ast \rangle_L$ and
D$_n$(4000) are only the rough age indicator of ULIRGs.

\subsection{The stellar mass}

ULIRGs experience the continuing conversion of gas to stars.  The
stellar masses in ULIRGs are mostly contributed by intermediate-age
and old stars (see $\mu_j$ distribution in Figure~\ref{2examp}), and
thus are a possible indicator for the ULIRG evolution.

The stellar mass of a ULIRG in the central 3$^{\prime\prime}$ for the
SDSS fiber spectrograph, $\mathcal{M}_{fiber}$, in unit of $M_{\odot}$,
can be estimated from the the stellar mass parameter
$\mathcal{M}_{cor\_tot}$ given by
STARLIGHT\footnote{http://www.starlight.ufsc.br/papers/Manual\_StCv04.pdf}:
\begin{equation}
\mathcal{M}_{fiber} =\mathcal{M}_{cor\_tot}\times
 10^{-17} \times4 \pi d^2\times(3.826\times10^{33})^{-1},
\end{equation}
where $d$ is the luminosity distance in cm. The aperture effect must
be corrected by using the SDSS Petrosian magnitude \citep{bde01},
$m_{\rm Petro}$, and fiber magnitude, $m_{\rm fiber}$, so that the estimated
mass can be a constant fraction of the total mass of galaxies,
independent of their positions and redshift. The aperture correction
factor for each SDSS photometric band is roughly given by \citep{hmn+03}:
\begin{equation}
A=10^{-0.4(m_{\rm Petro}-m_{\rm fiber})}.
\label{aper1}
\end{equation}
The spectra of ULIRGs fitted for the stellar population analysis
roughly cover the $u$, $g$, $r$, $i$ and $z$ bands. We take the
average of the correction factors of these bands weighted by
their uncertainty $\delta_A$,
\begin{equation}
\overline{A}=\sum_{j=u,g,r,i,z}\frac{A_j}{\delta_{A_{j}}^2}
/\sum_{j=u,g,r,i,z}\frac{1}{\delta _{A_{j}}^2},
\end{equation}
to correct the aperture effect for each ULIRG. The corrected stellar
mass, $\mathcal{M}_{*}$, in unit of $M_{\odot}$, which is a constant
fraction of the total stellar mass in a ULIRG, then can be estimated
by
\begin{equation}
\log(\mathcal{M}_{*})=\log(\mathcal{M}_{fiber})+\log(\overline{A}).
\end{equation}

\begin{figure}[tb]
\centering
\includegraphics[angle=270,width=75mm]{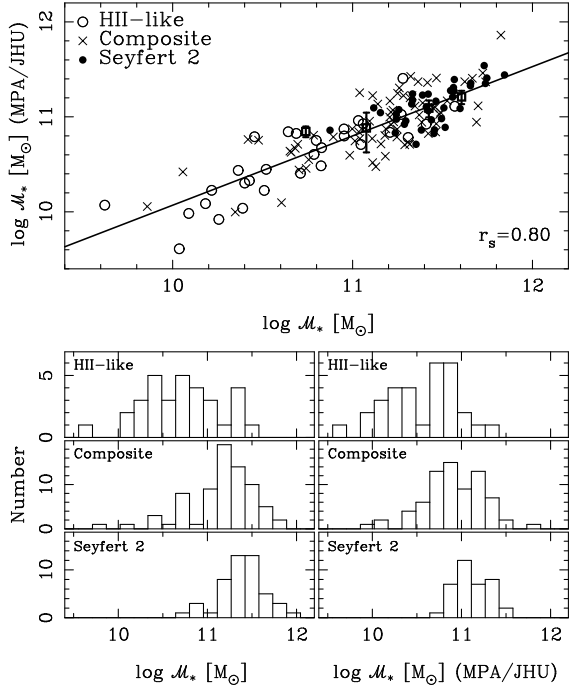}
\caption{Aperture-corrected stellar masses, $\mathcal{M}_{*}$, of
  ULIRGs calculated from the STARLIGHT parameters are compared with
  the stellar masses estimated from the SDSS photometric data by the
  MPA/JHU group (webpage in footnote No.5). The solid line is the best
  fit for all ULIRGs. The {\it total dynamical masses} of four ULIRGs,
  F09039+0503, F13428+5608, F15250+3609 and F15327+2340, are estimated
  from the central velocity dispersion and plotted as squares with
  error-bar. The distributions of the stellar masses for each type of
  ULIRGs are shown in the lower panels.}
\label{mass}
\end{figure}

We obtained $\mathcal{M}_{*}$ for the ULIRGs in our sample. { To show
  the uncertainty of such corrected mass estimates from the fiber
  spectrum of the central part of ULIRGs, we} compared them with the
total stellar masses of whole galaxies estimated by the MPA/JHU group
from the SDSS photometric data of $u$, $g$, $r$, $i$ and $z$ bands
\citep[see the footnote No.5 on their webpage,][]{src+07}. We found
that they are well correlated (see Figure~\ref{mass}) with the
Spearman Rank-Order Correlation Coefficient $r_s=0.80$, which means
that the aperture-corrected stellar masses calculated from the
STARLIGHT parameters are statistically fine. The data scatter in
Figure~\ref{mass} roughly indicates the uncertainty of the mass
estimates. \footnote{In our sample, the {\it total dynamical masses}
  of four ULIRGs, F09039+0503, F13428+5608, F15250+3609 and
  F15327+2340, can be calculated from the central velocity dispersion
  \citep{dtd+062}. As suggested by the referee, we got $\log \mathcal
        {M}_{dyn} [M_\odot]$ of 10.9, 11.1, 10.9, and 11.2 and plotted
        them in the Figure~\ref{mass} for comparison with the {\it
          total stellar masses} derived by the stellar population
        analysis. Considering the typical gas content of ULIRGs of
        $\sim$~10$^{10}M_\odot$ (see Section 4.3), we think different
        kinds of mass estimates are consistent with each other.}

The most important {feature} in Figure~\ref{mass} {is} the obviously
different distributions of $\log \mathcal{M}_{*}$ for HII-like,
composite and Seyfert 2 ULIRGs, whichever the masses estimated from
either spectral or photometric analysis are used. The mean of the
$\log \mathcal{M}_{*}$ distribution for the mass from the spectral
analysis for three types of ULIRGs, HII-like, composite, and Seyfert
2, are 10.70, 11.17 and 11.40, with the standard deviations of 0.45,
0.38 and 0.25, respectively. Seyfert 2 ULIRGs always have
$\mathcal{M}_{*} \gtrsim 10^{10.6}\; M_\odot$, larger than the masses
in HII-like or composite ULIRGs (see Figure~\ref{mass}). The increase
of the stellar masses along the sequence from HII-like ULIRGs to the
composite ULIRGs and Seyfert 2 ULIRGs is consistent with the standard
evolutionary scenario of ULIRGs.

\subsection{Gas in ULIRGs}

We noticed that Seyfert 2 ULIRGs mostly have $\mathcal{M}_{*}>
10^{10.6}M_\odot$ (see Figure~\ref{mass}). However, many HII-like
ULIRGs or composite ULIRGs have much smaller stellar mass,
$\mathcal{M}_{*} < 10^{10.5}M_\odot$. If the evolutionary sequence
from HII-like ULIRGs to composite ULIRGs and Seyfert ULIRGs is true,
there should be enough gas to be converted to stars. The gas mass in
ULIRGs is a key parameter to verify the standard evolution scenario of
ULIRGs.
\begin{figure}[tb]
\centering \includegraphics[angle=270,width=70mm]{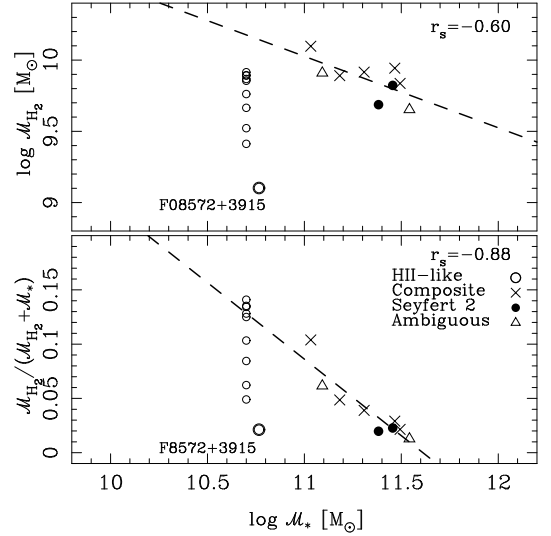}
\caption{The estimated gas mass from the CO measurements (the upper
  panel) and the gas fraction of total mass (gas plus stars) in 10
  ULIRGs (the lower panel) are plotted against the aperture-corrected
  stellar mass $\mathcal{M}_{*}$. The dashed lines are the best
  fittings to the 9 ULIRGs except for the outlier, an HII-like ULIRG
  F08572+3915. We also plotted the gas masses of another 9 HII-like
  ULIRGs which have CO observations available in literature but no
  SDSS spectra for stellar population analysis and hence no stellar
  mass. The mean value for HII-like ULIRGs (
  $\log\mathcal{M}_{*}=10.70$, see Figure~\ref{mass}) is adopted for
  these 9 HII-like ULIRGs.}
\label{gas}
\end{figure}

We collected the CO observations of ULIRGs to study their gas content.
Up to now, $\sim$~60 ULIRGs and LIRGs have measurements of CO(1-0)
\citep{sss91,sdrb97,cny+09}. However, only 10 of them have the optical
spectra available from SDSS, and we have got the relevant mass
estimates. CO observations of 9 ULIRGs were observed by \citet{sdrb97}
using the IRAM 30m telescope, and 1 ULIRG by \citet{cny+09} using
FCRAO 14m telescope. The gas mass (in unit of of $M_\odot$) can be
estimated from the CO luminosity (in unit of K~km~s$^{-1}$~pc$^2$)
\citep{ds98},
\begin{equation}
\mathcal{M}_{H_2} \approx 0.8\times L_{CO}.
\end{equation}
After the CO luminosity is converted to the same cosmology adopted in
this paper, we got the gas masses for these 10 ULIRGs (see
Figure~\ref{gas}). They are between $10^{9.6}-10^{10.2}M_\odot$,
typical for normal gas-rich galaxies, except for the only one HII-like
ULIRG, F08572+3915, which have a much smaller CO luminosity
\citep{sdrb97,cny+09} and hence the smaller gas mass. Clearly, more
gas exists in ULIRGs with a smaller stellar mass. We further collected
CO measurements \citep{sss91,sdrb97,cny+09} of 9 HII-like ULIRGs
\citep{vks99}, which have no SDSS spectra available for stellar
population analysis above. We take the mean $\log \mathcal{M}_{*}$ for
these HII-like ULIRGs, and plotted them in Figure~\ref{gas}.

The gas mass from the CO measurements (the upper panel of
Figure~\ref{gas}) and the gas fraction of total mass (the lower panel
of Figure~\ref{gas}) {  seem} to be anti-correlated with the stellar mass
for the ULIRGs ($r_s=-0.60$ and $r_s=-0.88$), if we do not consider
the outlier F08572+3915. The anti-correlation is only marginally
significant because of the small sample of data available. The
possible anti-correlation is preserved for the gas fraction and the
stellar mass estimated from photometric analysis.

HII-like ULIRGs with a gas fraction $>10\%$ follow the
anti-correlation. Gas in ULIRGs may be eventually converted to
stars. We noticed that about half of HII-like ULIRGs have a stellar
mass less than $10^{10.6}M_\odot$ (see Figure~\ref{mass}). HII-like
ULIRGs in Figure~\ref{gas} have a gas mass less than
$\sim10^{10}M_\odot$ and a gas fraction less than $\sim15\%$. For
those less massive HII-like ULIRGs, there is not enough gas to form
stars, and they are not massive enough in total to evolve to Seyfert 2
ULIRGs ($\mathcal{M}_{*}\sim10^{11}M_\odot$). Only massive HII-like
ULIRGs ($\mathcal{M}_{*}>10^{10.6}M_\odot$) may follow the standard
evolutionary scenario of ULIRGs and have a possible connection between
the starbursts and AGNs.

\section{Discussions}

Stellar population analysis for narrow-line ULIRGs show the
systematical changes of stellar age and mass along the sequence
from the HII-like ULIRGs, to the composite ULIRGs and Seyfert 2
ULIRGs. Here we discuss {several} factors which may affect the results
of stellar population analysis.

\subsection{Power-law AGN contribution for spectral continuum}

During the stellar population analysis with STARLIGHT, we used a
power-law component, $F(\lambda)\propto\lambda^{\alpha}$, with a
fixed $\alpha=-1.5$ to represent the AGN contribution to the
continuum spectrum \citep{cgm+04,cms+05}. The AGN contribution is
significant for composite, Seyfert 2 and LINER ULIRGs, but not for
HII-like ULIGRs. However, the index for the power-law contribution,
$\alpha$, could be $-2.0$ to $-1.0$ \citep[see][]{ngcl98}.  To
evaluate the influence of the power-law index $\alpha$ in the
stellar population analysis, we re-do the fittings to the all
spectra of 160 ULIRG sample with different values of $\alpha$, from
$-1.0$, $-$1.25, $-$1.5, $-$1.75, to $-$2.0, for ten times each with
different random seeds. We found that the change of $\chi^2$ by
different $\alpha$ values is less than $\sim $1\%, which is not
adequate to use $\chi^2$ to judge the best $\alpha$ for fitting. {
Changes of the stellar mass
  and the mean stellar age caused by different $\alpha$ are always
  less than 10\%.} Therefore, the adopted power-law index of
$\alpha=-1.5$ is fine for the stellar population analysis.

\subsection{Different bases of simple stellar populations}

The evolutionary population synthesis method is mainly based on the
stellar evolution model. We have used a base of 45 simple stellar
populations (SSPs) for the stellar population analysis. The
different bases of SSPs may affect the fitting results. Following
the same procedures described in Section 3, we re-do the fittings to
the spectra for 160 ULIRGs with a base of 150 SSPs (25 ages and 6
metallicities) and a base of 15 SSPs (15 ages and 1 metallicity)
extracted from \citet{bc03}.  {Although} the fractional contribution
of each stellar population component does not remain the same and
may vary about 20\%, the mean stellar age and stellar mass do not
change {  significantly, less than $\sim$10\%. Their distributions} are
consistent with those shown in Section 4 for different types of
ULIRGs.

\begin{figure}[tb]
\centering \includegraphics[angle=270,width=83mm]{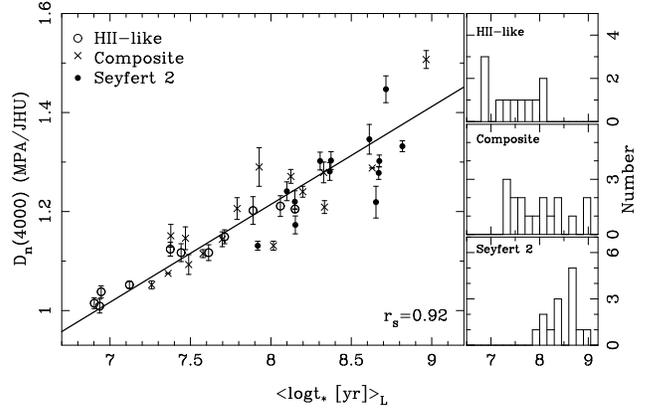}
\caption{Same as Figure~\ref{age} but only for ULIRGs in the
redshift range of 0.20$-$0.24.}
\label{age_z}
\end{figure}

\subsection{The aperture effect}

We made stellar population analysis for the SDSS spectrum of the
central $3''$ region. The aperture effect has been corrected for the
mass estimates, but not for the mean stellar age.  To evaluate the
aperture effect, we divided the ULIRG sample to several redshift
bins, and re-do the analysis. In each redshift bin, the linear size
of the aperture is almost the same for all ULIRGs, and the aperture
effect is almost the same.  Here we take the redshift bin of
$0.20<z<0.24$ as the example which has the largest number of ULIRGs
in the bin.  As shown in Figure~\ref{age_z}, the correlation between
$D_n$(4000) and $\langle log t_\ast \rangle_L$ keeps the same as the
whole sample of ULIRGs. The distributions of $\langle log t_\ast
\rangle_L$ for HII-like, composite and Seyfert 2 ULIRGs in the right
panel of Figure~\ref{age_z} also show {  similar offsets as the
whole
  sample. So do for other redshift bins.} Therefore, we believe that
the aperture effect does not influence our results.

\subsection{Extinction}

In the stellar population synthesis model of STARLIGHT
\citep{cms+05,msc+06}, the extinction was considered as a foreground
screen and fitted by one parameter, the $V$-band extinction $A_V$. But
the young, intermediate and old stellar populations have different
extinctions in a galaxy. Particularly the young stellar population was
more extinguished than the others, being in places very dust obscured.
Different $A_V$ should be assumed for the different populations.
However, this can not be achieved in the model.

\begin{figure}[bt]
\centering \includegraphics[angle=270,width=85mm]{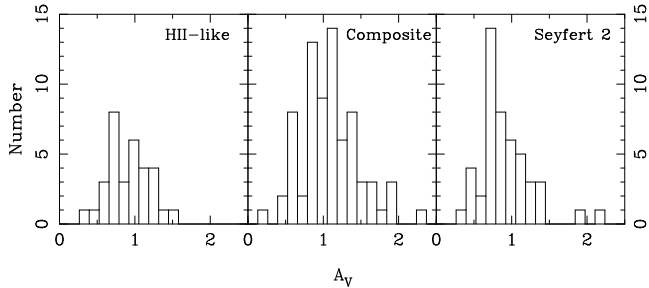}
\caption{Distribution of the extinction parameter $A_V$ given by
  the STARLIGHT model for HII-like ULIRGs, composite ULIRGs and
  Seyfert 2 ULIRGs.}
\label{extinc}
\end{figure}

From the stellar population analysis to ULIRGs, the distributions of
extinction parameter $A_V$ given by the STARLIGHT model for HII-like
ULIRGs, composite ULIRGs and Seyfert 2 ULIRGs are very similar, as
shown in Figure~\ref{extinc}. The means of $A_V$ for three types of
ULIRGs are 0.91, 1.09 and 0.92 with the standard deviations of 0.28,
0.39 and 0.36, respectively. The extinction parameters $A_V$ of
HII-like and Seyfert 2 ULIRGs on average are {  slightly} smaller
than that of composite ULIRGs, which is consistent with
\citet{vrk+09}.

The distributions of the mean stellar age for different ULIRGs
discussed in Section~4 are probably not influenced by the
simplified treatment of extinction in the model. The mean stellar age
given by the stellar population synthesis model are well correlated
with results of D$_n$(4000) (see Figure~\ref{age}), which verifies the
mean age. On the other hand, in ULIRGs, the stellar mass is dominant
by old stellar populations (see the right panels of
Figure~\ref{2examp}), which are expected to suffer less from
extinction than young stellar populations. Therefore, the simplified
extinction in the model probably does not affect our conclusions for
ULIRGs.

\section{Conclusions}

We analyse the stellar populations for a sample of 160 narrow-line
ULIRGs. They are optically classified as 32 HII-like ULIRGs, 74
Seyfert-HII composite ULIRGs, 6 LINER ULIRGs and 48 Seyfert 2
ULIRGs. We found that that along the sequence of HII-like, composite
and Seyfert 2 ULIRGs, both the mean stellar age and the
aperture-corrected stellar mass increase. This supports the standard
evolutionary scenario of ULIRGs in which the Seyfert 2 ULIRGs are in a
late stage of ULIRG evolution with a large mean stellar age and a
large stellar mass ($\sim10^{11} M_{\odot}$) while HII-like ULIRGs are
in an early stage with a young stellar age and a small stellar mass on
average.

Whether HII-like ULIRGs have enough gas for starburst, so that they
can evolve to Seyfert 2 ULIRGs? We collected CO measurements of 10
ULIRGs which have the SDSS spectra for the stellar population
analysis. The gas mass fractions seem to be anti-correlated with the
stellar masses for massive ULIRGs. All ULIRGs in our sample with CO
measurements have a gas mass less than $\sim10^{10} M_{\odot}$.
HII-like ULIRGs with a small stellar mass ($\mathcal{M}_{*} <
10^{10.4}M_\odot$) do not possess enough gases for starburst, and
therefore have no evolution connections with massive Seyfert 2
ULIRGs. We conclude that only massive HII-like ULIRGs and
composite ULIRGs may follow the evolution sequence toward AGNs.

\bigskip

\begin{acknowledgments}
  {We thank the referee for helpful comments, and} Dr. XiaoYan Chen,
 XianMin Meng and Ran Wang for discussions and suggestions.
  The authors are supported by the National Natural Science Foundation
  of China (10821061, 10833003 and 11033001), and the National Key
  Basic Research Science Foundation of China (2007CB815403,
  2007CB815405).
  The STARLIGHT project is supported by the Brazilian agencies CNPq,
  CAPES and FAPESP and by the France-Brazil CAPES/Cofecub program.
  Funding for the SDSS has been provided by the Alfred P. Sloan
  Foundation, the Participating Institutions, the National Science
  Foundation, the Department of Energy, the National Aeronautics and
  Space Administration, the Japanese Monbukagakusho, the Max Planck
  Society, and the Higher Education Funding Council for England. The
  SDSS Web Site is http://www.sdss.org.
\end{acknowledgments}

\bibliographystyle{apj}
\bibliography{rev2}

\end{document}